\documentstyle[12pt,epsfig,rotating]{article}
\newcommand{\AmS}{{\protect\the\textfont2
  A\kern-.1667em\lower.5ex\hbox{M}\kern-.125emS}}
    \def\r#1{$^{[#1]}$}
\def\E{\mbox{e}^+\mbox{e}^-}

\def\ifmath#1{\relax\ifmmode #1\else $#1$\fi}%

\def\s2{\hskip2pt}  \def\d{{\rm d}}    \def\e{{\rm e}}
\def\f{\left}   \def\g{\right}
\newcommand{\beqa}{\begin{eqnarray}} \newcommand{\eeqa}{\end{eqnarray}  }
\newcommand{\beqan}{\begin{eqnarray*}} \newcommand{\eeqan}{\end{eqnarray*}}
\newcommand{\beq}{\begin{equation}} \newcommand{\eeq}{\end{equation}  }
\def\cl{\centerline} \def\bcc{\begin {center}} \def\ecc{\end {center}}
\def\btbl{\begin{tabular}} \def\etbl{\end{tabular}}
\def\CSFM{{\footnotesize CSFM}} \def\NUFM{{\footnotesize NUFM}}
\def\RQMD{{\footnotesize RQMD}} \def\QCD{{\footnotesize QCD}}
\def\CERN{{\footnotesize CERN}} \def\SPS{{\footnotesize SPS}}
\def\E877{{\footnotesize E877}} \def\QGP{{\footnotesize QGP}}
\def\BNL{{\footnotesize BNL}} \def\AGS{{\footnotesize AGS}}
\def\RHIC{{\footnotesize RHIC}} \def\LHC{{\footnotesize LHC}}

\begin{document}
\null{}\vskip -1.2cm
\hskip12cm{\bf HZPP-0004}
\vskip-0.2cm

\hskip12cm May. 5, 2000

\vskip1cm

\centerline{\Large  
Thermal Equilibrium and Non-uniform Longitudinal Flow}
\vskip0.3cm
\centerline{\Large in Relativistic Heavy Ion Collisions\footnote{Work
supported in part by the NSFC under project 19775018.}}
\vskip1cm
\centerline{\large FENG Shengqin, LIU Feng and LIU Lianshou}
\vskip0.4cm
\centerline{\small Institute of Particle Physics, Huazhong Normal University, 
Wuhan, 430079, China}

\vskip0.9cm

\vskip2cm
\begin{center}{\large ABSTRACT}\end{center}
\vskip0.5cm
\begin{center}\begin{minipage}{124mm}
{\small \hskip0.8cm
A model with non-uniform flow in the longitudinal direction is proposed
for the 
relativistic heavy-ion collisions and compared with the 14.6 A GeV/$c$
Si-Al and 10.8 A GeV/$c$ Au-Au collision data. 
The stronger influence of transparency on the distribution of heavier
produced particles and the larger stopping 
in the heavier collision system are accounted 
for by using a new geometrical parameterization picture.  
The central dips in the proton 
and deuteron rapidity distributions for Si-Al collision are reproduced.}
\end{minipage}\end{center}
\vskip3cm
PACS number(s): {25.75.-q }
\newpage

\noindent
{\bf I. Introduction}\\
\noindent

The experimental finding that colliding nuclei are not transparent but undergo 
a violent reaction in central collisions represents one of the major 
motivations for the study of ultra-relativistic heavy ion collisions at the 
\CERN/\SPS, \BNL/\AGS\ and also at the future \BNL/\RHIC\ and \CERN/\LHC. 
Of central 
importance is the ability of understanding to what extent the nuclear
matter has been compressed and heated.

The use of thermal models to interpret data on particle distribution from
nuclear collisions is motivated by the hope and expectation that in
collision between sufficiently large nuclei at sufficiently high
energies a state of exicted nuclear matter close to local thermodynamic
equilibrium can be formed, allowing us to study the thermodynamics of 
\QCD\ and the possible phase transition from a hadronic gas to a \QGP.
Since we do not yet reliably know how big the collision system and how 
large the beam energy have to be for this to occur (if at all), thermal 
models should be considered as an importent  phenomenological tool to test 
for such a behavior. 

The study of collective flow in high energy nuclear collisions has attracted 
increasing attentions from both experimental\r{1} and theoretical\r{2}
point of view. 
The rich physics of longitudinal and transverse flows is due to their
sensitivity to the system evolution at early time. The expansion and cooling 
of the heated and highly compressed matter could lead to a considerable 
collectivity in the final state. Due to the high pressure, particles might
be boosted in the transverse and longitudinal directions. The collective 
expansion of the system created during a heavy-ion collision implies 
space-momentum correlation in particle distributions at freeze-out. 

The experimental data on the rapidity distributions of produced particles
in 14.6 A GeV/$c$ Si-Al collisions has been ultilized to study the 
collective expansion using a cylindrical-symmetric longitudinal flow 
model (\CSFM)\r{3,4}. 
In this model, the distribution of final-state particles comes from the 
superposition of a number of fire-balls, distributed uniformly within 
the kinematical limit along the longitudinal (rapidity $y$) axis.   
The model-results fit well the experimental distribution of pion, 
but are too narrow in the case of heavier particles, such as proton 
and deuteron.  In particular, the central dip, which can be clearly 
seen in the distribution of proton, is not reproduced. 

More recently, \E877\ Collaboration\r{5} has published their data for 
10.8 A GeV/$c$ Au-Au collisions, which provides a good chance to compare 
the stopping power in collision systems of different sizes. 
A possible central peak of the rapidity distribution of proton at
around mid-rapidity, which was obtained through extrapolating the 
experimental data to mid-rapidity using \RQMD\ model\r{6}, has been taken as an 
evidence for the increasing of stopping power, but the reliability of this
extrapolation is model-dependent.

Stopping and transparency describe the same physical aspect in relativistic 
heavy ion collision from two opposite sides. In order to settle a reliable
model for relativistic heavy ion collision, this aspect must be taken into 
account carefully. The assumption of uniformly distributed fire-balls in 
the \CSFM\ model is a crude one. It does not account for the memory of the 
fire-balls on the motion of the incident nuclei. In the present paper we 
propose a non-uniform longitudinal flow model (\NUFM) using 
a new geometrical parameterization picture\r{7}
to describe the nuclear stopping/transparency in a more proper way. 
The stronger influence of transparency on the distribution of heavier
produced particles (proton and deuteron) as well as the larger stopping 
in the heavier collision system (Au-Au) are described by using this picture. 
The central dips in the 
proton and deuteron rapidity distributions for Si-Al collisions are 
reproduced, and at the same time the central peak in the pion rapidity
distributions is maintained.

In section II the non-uniform longitudinal flow model (\NUFM) with 
a geometrical parameterization picture  is 
formulated. The results of the model are given and compared with
the experimental data in section III.  A short summary and conclusion is 
given in section IV. In order to avoid the complexity associated  
with the production
of strange particles and concentrate on the expansion of the system, we will 
discuss in this paper only normal non-strange particles ------ pions, protons 
and deuterons.

\vskip1.0cm
\vskip0.2cm
\noindent
{\bf II. Non-uniform longitudinal flow } \\
\vskip0.2cm

Firstly, let us briefly recall the fireball scenario of relativistic heavy ion 
collisions. 

Since the temperature at freeze-out exceeds 100 MeV, the 
Boltzmann approximation is used. Transformed into rapidity $y$ and
transverse momentum $p_{t}$ this implies\r{4}:

\begin{equation}  
  E\frac {\d^{3}n}{\d^{3}p} \propto E\e^{(-E/T)}=    
  m_{t}\cosh y \ \e^{(-m_{t}\cosh y/T)} 
\end{equation}
\noindent 
Here $m_{t}=\sqrt{m^{2}+p_{t}^{2}}$ is the transverse mass, $m$ is the mass 
of the produced particles at freeze-out.

The rapidity is defined as $y=\tanh^{-1}(p_{l}/E)$, where $p_{l}$ is the 
longitudinal momentum of the produced particle. Substituting into Eqn.(1) 
and integrating over $m_{t}$, we get the rapidity distribution of the 
isotropic thermal source,
\begin{equation}  
\frac {\d n_{\rm iso}}{\d y} \propto 
\frac {m^{2}T}{(2\pi)^{2}}(1+2\xi_0+2\xi_0^{2})\e^{(-1/\xi_0)}.  
\end{equation}
\noindent
where $\xi_0={T}/{m\cosh y}$. 

Eqn's.1 and 2 give the isotropic thermal distribution.  
As mentioned in Ref.[4], for pions with 
$m\approx{T}$, this distribution is close to that of massless particles, i.e.
proportional to $1/\cosh^{2} y$. For heavier particles isotropic emission 
implies a strongly narrow distribution, in contradiction to the experimental
findings, see e.g. the dashed lines in Fig.6. 

The measured momentum distribution of the final-state particles is certainly 
aniso-tropic. It is privileged in the direction of the incident nuclei. 
This is because the produced hadrons still carry
their parent's kinematic information, making the longitudinal 
direction more populated than the transverse ones.
The simplest way\r{3,4} to account for this anisotropy is to add up the 
contributions from a set of fire-balls 
with centers located uniformly in the rapidity region 
[$-y_0, y_0$], as sketched schematically in Fig.1. The corresponding 
rapidity distribution is obtained through 
changing the $\xi_0$ in Eqn.(2) into $\xi={T}/{m\cosh(y-y')}$ and integrating 
over $y'$ from $-y_0$ to $y_0$:
\begin{equation}  
\frac {\d n_{_{\rm CSFM}}}{\d y} \propto 
\int_{-y_0}^{y_0} \d y' 
\frac {m^{2}T}{(2\pi)^{2}}(1+2\xi+2\xi^{2})\e^{(-1/\xi)},   
\end{equation}
where $\xi={T}/{m\cosh(y-y')}$. 
Equivalently, we can also use the angular variable 
$\Theta$ defined by $\Theta = 2 \tan^{-1} \exp (-y')$,    
and change the integration variable in Eqn.(3) to $\Theta$, 
\beqa   
\frac {\d n_{_{\rm CSFM}}}{\d y} &=& eKm^2T 
\int_{\Theta_{\rm min}}^{\Theta_{\rm max}} \frac{\d \Theta}{\sin \Theta} 
\f(1+\frac{2T}{m\cosh(y-y')}+ \frac{2T^2}{m^2\cosh^2(y-y')} \g)
\nonumber \\
&\null{}& \hskip3cm \times
\exp(-m\cosh(y-y') / T)\ ,  
\eeqa
cf. the solid circle and lines in Fig.2. 

This simple approach fits the rapidity distribution of pions 
well but
fails to reproduce the central dip in heavier produced particles, which
is clearly seen in the experimental distributions of protons and deuterons. 

Note that in this \CSFM\ model the 
distribution of fire-balls in the longitudinal direction of phase space 
is uniform which 
does not account for the interaction of the incident nuclei when they pass 
through each other properly. This is a crude approximation.  
A more reasonable assumption is that the fire-balls keep some memory
on the motion of the incident nuclei, and therefore
the distribution of fire-balls, instead of being uniform in the 
longitudinal direction, is more concentrated in the direction of motion of
the incident nuclei, i.e. more dense at large absolute value of rapidity. 
A parametrization for such a distribution can be obtained by using 
an ellipse like picture on emission angle distribution, as
shown in Fig.3.  In this scenario, the emission angle is 
\beq   
\theta = \tan^{-1} (e \tan\Theta),
\eeq
\noindent
where the parameter $e (0\leq e \leq 1$)
represents the ellipticity of the introduced ellipse which describes
the non-uniformity of fire-ball distribution in the longitudinal direction,
as sketched in Fig.4.

Subsituting Eqn.(5) together with $y_{\rm e}=-\ln\s2\tan(\theta/2)$ into 
Eqn.(3), a distribution function $Q(\theta)$ for the  emission angle of 
non-uniform flow is introduced in the integration
\beqa   
\frac {\d n_{_{\rm NUFM}}}{\d y} &=& eKm^2T 
\int_{\theta_{\rm min}}^{\theta_{\rm max}}  
\frac{Q(\theta) \d \theta}{\sin \theta}
\f(1+\frac{2T}{m\cosh(y-y_{\rm e})}+ \frac{2T^2}{m^2\cosh^2(y-y_{\rm e})} \g) 
\nonumber \\
&\null{}& \hskip3cm \times
\exp(-m\cosh(y-y_{\rm e}) / T) ,    
\eeqa
\beq 
y_{\rm e}=-\ln\s2\tan(\theta/2)\ , \qquad 
Q(\theta) = \frac{1}{\sqrt{e^2+\tan^2\theta} |\cos\theta| }.
\eeq
\noindent
Here $\theta_{\rm min}=2\tan^{-1}(e^{-y_{\rm e0}})$, 
$\theta_{\rm max}=2\tan^{-1}(e^{y_{\rm e0}})$.
$y_{\rm e0}$ is the rapidity limit which confines the 
rapidity interval of longitudinal flow. 

Changing the integration variable in Eqn.(6) back to
$y_{\rm e}$, the rapidity distribution can be rewritten as follows: 
\beqa   
\frac {\d n_{_{\rm NUFM}}}{\d y} &=& eKm^2T 
\int_{-y_{\rm e0}}^{y_{\rm e0}} 
\f(1+\frac{2T}{m\cosh(y-y_{\rm e})}+ \frac{2T^2}{m^2\cosh^2(y-y_{\rm e})} \g) 
\nonumber \\
&\null{}& \hskip3cm \times
\exp(-m\cosh(y-y_{\rm e}) / T)    
\rho(y_{\rm e}) \d y_{\rm e}.
\eeqa
\beq 
\rho(y_{\rm e}) = \sqrt{\frac {1+\sinh^{2}(y_{\rm e})}{1+{\rm e}^{2}\sinh^{2}(y_{\rm e})}}.
\eeq
\noindent
Here $\rho(y_{\rm e})$ is the distribution function of non-uniform 
flow in the 
longitudinal direction. It can be seen from fig.5, that the larger is the 
parameter $e$, the flatter is the distribution $\rho(y_{\rm e})$ and
the more uniform is the longitudinal-flow distribution. 
When $e\rightarrow{1}$, 
the longitudinal-flow distribution is completely uniform 
($\rho(y_{\rm e})\rightarrow{1}$), returning back to the \CSFM\ model.

\vskip1.2cm
\noindent
{\bf III. Comparison with experiments  } \\
\vskip0.2cm

We now proceed to compare the model-results with the \AGS\ data. 

The rapidity distributions of pion, proton and deuteron 
for 14.6 A GeV/$c$ Si-Al collisions{\r{9,10}} 
are given in Fig.6 ($a, b$ and $c)$. 
The dashed, dotted and solid lines (band) correspond to the results from 
isotropical thermal model, uniform longitudinal 
flow model (\CSFM) and the non-uniform longitudinal flow model (\NUFM) 
respectively. The rapidity limit $y_{\rm e0}$ and the ellipticity $e$
used in the calculation are listed in Table I
and illustrated in Fig.7. The  rapidity limit 
$y_{\rm 0}$ used in the \CSFM\ model of
Ref. [4] is also listed in Table I for comparison.
The parameter $T$ is chosen to be 0.12 GeV following Ref.[4].

Note that the distribution of the light particle (pion) is insensitive to the 
ellipticity $e$. Changing $e$ from 0.35 to 0.7 results in a narrow band shown
in Fig.6($a$). Since pions are produced through the interaction of colliding
nuclei, they have less memory on the motion of the incident nuclei before
interaction. Therefore, physically the value of ellipticity $e$ for pion
should be bigger than that for proton and deuteron, but the exact value
cannot be fixed through fitting the model results to the experimental data.

\vskip0.5cm
\cl{Table I \ \ The value of model-parameters}
\vskip-0.5cm
\bcc\btbl{|c|c|c|c|c|c|}\hline
           & \multicolumn{3}{|c|}{Si-Al Collisions}
        &\multicolumn{2}{|c|}{Au-Au Collisions} \\ \cline{2-6}
 Parameter & $\pi$ & p & d  & $\pi$ & p  
\\ \hline
$e$ & 0.35 -- 0.7 & 0.52 & 0.56 & 0.35 -- 0.7 & 0.58 
\\ \hline
$y_{\rm e0}$ & 1.35 & 1.35 & 1.35 & 1.05& 1.05  
\\ \hline
$y_{\rm 0}$ & 1.15 & 1.15 & 1.15 & &   \\
\hline
\etbl\ecc

It can be seen from the figures that
the \NUFM\ model reproduces the central dip of the rapidity distribution 
of heavier particles  (proton and deuteron) in agreement with 
the experimental findings, while for light particles (pions) there is 
a peak instead of dip at central rapidity. 
Note that the appearance or disappearance of central dip is insensitive to
the rapidity limit $y_{\rm e0}$ but depends strongly on the magnitude 
of the ellipticity $e$ and the mass $m$ of the produced particles.  
For the heavier particles (proton and deuteron) a central dip appears for
$e < 0.8$, but for light particles  (pions) there is no dip even when $e$ 
is as small as 0.35, cf. Table I and Fig. 6.

On the other hand, the width of the rapidity distributions is mainly
controlled by the amplitude $y_{\rm e0}$ of the longitudinal flow. 
A single value (1.35) of $y_{\rm e0}$ can account for the wide distribution 
of heavier particles (protons and deuterons) and at the same time fit the 
pion-distribution well. The difference in the width of the $\d N/\d y$
distribution originates mainly from the difference in the mass of the 
particles\r{3,4}.

In Fig.8 are shown the rapidity distributions of pions and protons 
for Au+Au collisions at 10.8 A GeV/$c${\r 5}.
The dotted and solid lines (band) correspond to
the results of \CSFM\ and \NUFM\ models (with parameters listed in Table I)
respectively. In the calculation the \CSFM\ results
we have also used the \NUFM\ model with the same rapidity limit  
$y_{\rm e0}$ listed in Table I but with ellipticity $e=1$.
The histogram is the result from the \RQMD\ model.

It can be seen from Fig.8 that in the \NUFM\ model there is a shallow dip
in the central rapidity of the distribution of proton, instead of 
a central peak as predicted by the \CSFM\ model. 
However, the presently available experimental data are restricted to the large
rapidity. The peak at central rapidity is the extrapolation of data using
\RQMD\ and is model dependent. 
It is interesting to see whether the prediction of a central dip
(plateau) or a central peak will be observed in future experiments.

Comparing the parameter values for Si-Al (smaller colliding nuclei) and 
Au-Au (larger colliding nuclei) collisions listed in Table I, it can be seen 
that the rapidity limit $y_{\rm e0}$ is smaller and the ellipticity $e$ 
for proton is 
bigger for the larger colliding nuclei than for the smaller ones. Both of
these two show that the hadronic system formed from the larger colliding 
nuclei is less spread out in rapidity, 
i.e. there is stronger nuclear stopping in the collision of larger nuclei.

\vskip1.2cm
\noindent
{\bf IV. Summary and Conclusions}
\vskip0.2cm

In high energy heavy-ion collisions, due to the transparency of nucleus 
the participants
will not lose the historical memory  and the produced hadrons will 
carry some of their 
parent's memory of motion, leading to the unequivalence in longitudinal
and transverse directions. So it is reasonable to assume that the flow of 
produced particles is privileged
in the longitudinal direction. This picture has been used by lots of 
models\r{3,8}. Here we would mention two 
thermal and hydrodynamic models, one is the
boost-invariant longitudinal expansion model postulated by Bjorken \r{8}
which can explain such an anisotropy already at the level of 
particle production in hadron-hadron collisions.  This model has been
formulated for asymptotically high energies, where the rapidity 
distribution of
produced particles establishes a plateau at midrapidity. The second model 
is the 
\CSFM\ postulated first by Schnedermann, 
Sollfrank and Heinz\r{3} which
accounts for the anisotropy of longitudinal and transverse direction 
by adding the contribution from
a set of fire-balls with centers located uniformly in the rapidity 
region [-$y_{\rm 0}$,$y_{\rm 0}$] in the longitudinal direction,
sketched schematically in Fig's.1 and 2 as dashed circles. 
It can account for the wider rapidity distribution
when comparing to  the prediction of the pure thermal 
isotropic model but fails to reproduce
the central dip in the proton and deuteron rapidity distributions.  

In this paper, we argue that the transparency/stopping of relativistic heavy 
ion collisions should be taken into account more carefully. It will not
only lead to the anisotropy in longitudinal-transverse directions, but 
also render the fire-balls (especially for those of proton and deuteron)
concentrate more in the direction of motion of the incident nuclei. 
A non-uniform longitudinal flow model is proposed, 
which assumes that the centers of fire-balls are distributed non-uniformly
in the longitudinal phase space.  A parameter $e$ is introduced through a
geometrical parametrization which can express the non-uniformity
of flow in the longitudinal direction, i.e.
the centers of fire-balls of proton and deuteron 
prefer to accumulate in the two extreme
rapidity regions ($y_{\rm e} \approx \pm y_{\rm e0}$)
in the c.m.s. frame of relativistic heavy-ion collisions, 
and accordingly the distribution is diluted 
in the central rapidity region ($y_{\rm e} \approx 0$).

It is found that the depth of the central dip  
depends on the magnitude of the parameter $e$ and 
the mass of produced particles, i.e. the non-uniformity of longitudinal flow
which is described by the parameter $e$ determines the depth of the central 
dip for heavier particles. On the other hand, 
the central peak in the pion distribution 
turns out to be insensitive to the value of $e$ and can be 
well reproduced from this model simultaneously with the
central dip in the proton and deuteron rapidity distributions.

Through comparing the feature of collision systems of different size, 
it is found that the maximum flow velocities are smaller for 
the heavier collision system than for the lighter 
ones, which suggests, together with the larger $e$,
a larger stopping in the bigger collision system. 

\vskip 1.0cm

 \newpage
\noindent{\bf\large\bf Figure captions}

\vskip0.5cm
\noindent {\large Fig.1} \ Schematic sketch of the distribution of
fire-balls in the uniform flow model (\CSFM).

\vskip0.5cm
\noindent {\large Fig.2} \ Schematic sketch of the emission angle
$\Theta$ (solid circle and lines) and the corresponding distribution
of fire-balls (dashed circles) in the uniform flow model (\CSFM). 

\vskip0.5cm
\noindent {\large Fig.3} \ Schematic sketch of the  emission angle
$\theta$ in the non-uniform flow model (\NUFM).

\vskip0.5cm
\noindent {\large Fig.4} \ Schematic sketch of the distribution of
fire-balls in the non-uniform flow model (\NUFM).

\vskip0.5cm
\noindent {\large Fig.5} \ The distributions $\rho(y_{\rm e})$ 
of the center of fire-balls as a function of $y_{\rm e}$. 
When $e\rightarrow 1$, the non-uniform distribution turns to the
uniform distribution $\rho(y_{\rm e})\rightarrow 1$.

\vskip0.5cm
\noindent {\large Fig.6}\ Rapidity distributions for central 14.6 A GeV/$c$
Si+Al collisions. Open circles are the experimental data for Si+Al collisions
taken from Ref.'s [9,10]. 
Dashed, dotted and solid lines are the distributions from the 
isotropical thermal model, cylindrical-symmetric longitudinal flow model 
(\CSFM) and non-uniform longitudinal flow model (\NUFM) respectively. 
Fig's. ($a$), ($b$) and ($c$) are the pion, 
proton and deuteron distributions respectively. 
The ellipticity $e$ for pion is within a range [0.35, 0.7] and therefore
the \NUFM\ results are present as a band. The temperature $T=0.12$ GeV. 

\vskip0.5cm
\noindent
{\large Fig.7} \  The fire-ball distribution functions 
$\rho(y_{\rm e})$ verus rapidity $y_{\rm e}$
in the  non-uniform flow of different particles for 
the collisions of Si+Al at 14.6 A GeV/$c$ 
($a$) and Au+Au at 10.8 A GeV/$c$ ($b$) respectively.
For pion the  distribution functions $\rho(y_{\rm e})$
are only ploted for the two boundaries of the 
used region $0.35 \leq e \leq 0.7$.

\vskip0.5cm
\noindent
{\large Fig.8} Rapidity distributions for pions ($a$) and protons ($b$) in 
central Au+Au collisions at 10.8  A GeV/$c$. Full circles 
represent the experimental data taken from Ref.[5].
Open circles are the reflection of the data. 
The solid  line is our calculation using the \NUFM\ model. 
The ellipticity $e$ for pion is within a range [0.35, 0.7] and therefore
the results for pion are present as a band. 
The histogram shows the results from \RQMD\ and the dotted lines
are the results from the prediction of \CSFM\ model. The temperature 
$T=0.14$ GeV. 
  
\newpage
\begin{center}
\begin{picture}(250,450)
\put(-40,340)
{
{\epsfig{file=fig1.epsi,width=320pt,height=54pt}}
}
\put(-40,80)
{
{\epsfig{file=fig2.epsi,width=350pt,height=150pt}}
}
\end{picture}
\end{center}

\vskip-12cm
\cl{Fig. 1}

\vskip12cm
\cl{Fig. 2}

\newpage
\begin{center}
\begin{picture}(250,450)
\put(30,300)
{
{\epsfig{file=fig3.epsi,width=220pt,height=170pt}}
}
\put(-40,100)
{
{\epsfig{file=fig4.epsi,width=320pt,height=54pt}}
}
\end{picture}
\end{center}
\vskip-10cm
\cl{Fig. 3}
\vskip8cm
\cl{Fig. 4}

\newpage
\begin{center}
\begin{picture}(250,450) 
\put(25,300)
{
{\epsfig{file=fig5.epsi,width=180pt,height=180pt}}
}
\put(25,-100)
{
{\epsfig{file=fig6.epsi,width=180pt,height=280pt}}
}
\end{picture}
\end{center}

\vskip-10cm
\cl{Fig. 5}

\vskip14cm

\cl{Fig. 6}

\newpage
\begin{center}
\begin{picture}(250,450) 
\put(-15,100)
{
{\epsfig{file=fig7.epsi,width=240pt,height=380pt}}
}
\end{picture}
\end{center}

\vskip-1cm
\cl{Fig. 7}

\newpage
\begin{center} 
\begin{picture}(250,450)
\put(0,200)
{
{\epsfig{file=fig8.epsi,width=220pt,height=240pt}}
}
\end{picture}
\end{center}
\vskip-6cm
\cl{Fig. 8}

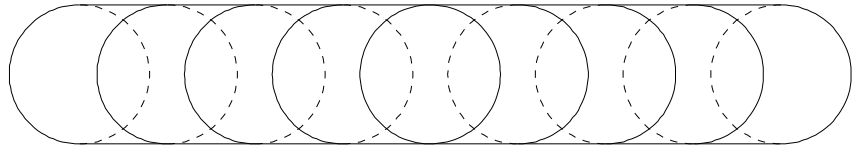
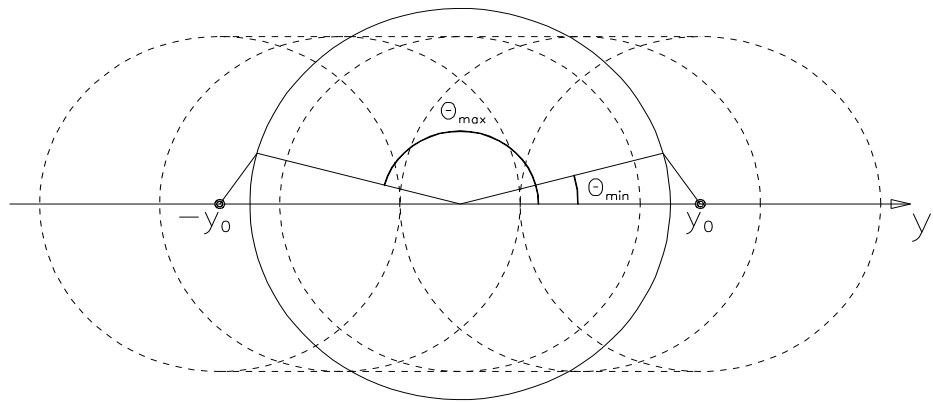
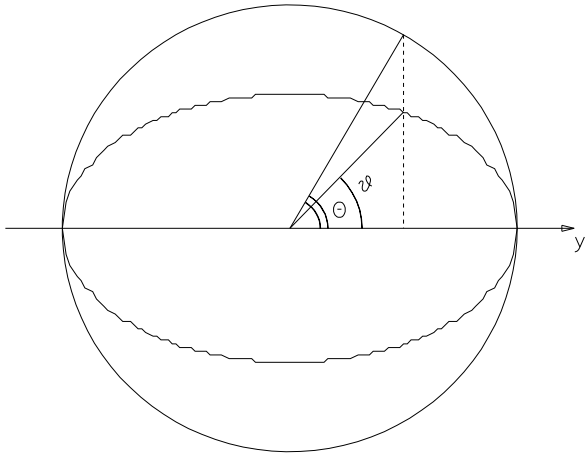
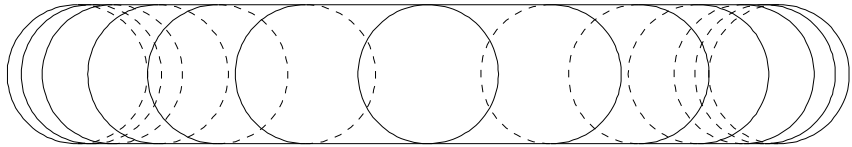
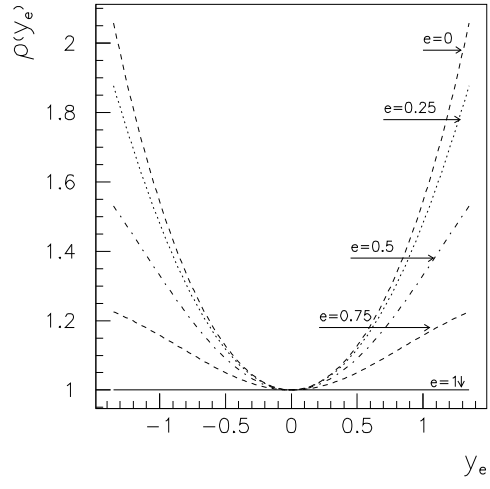
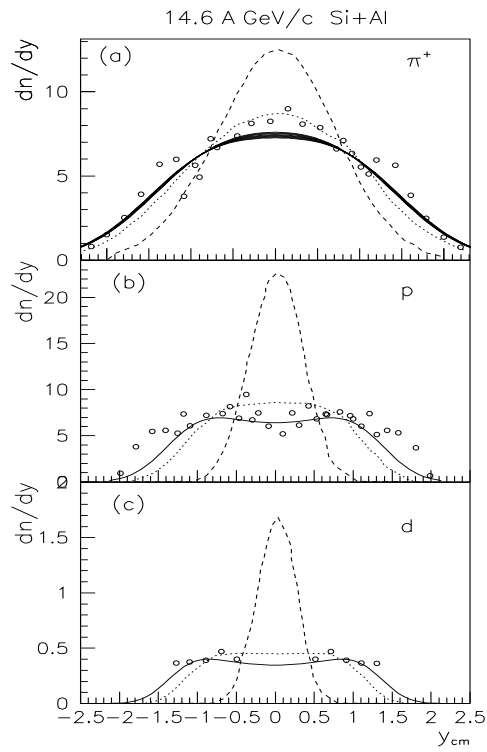
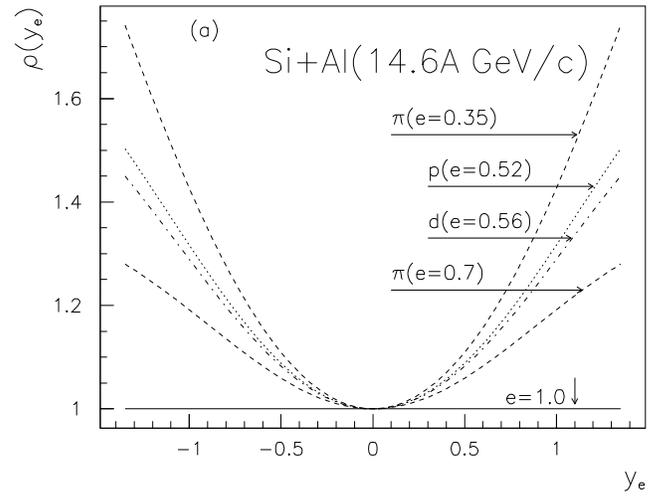
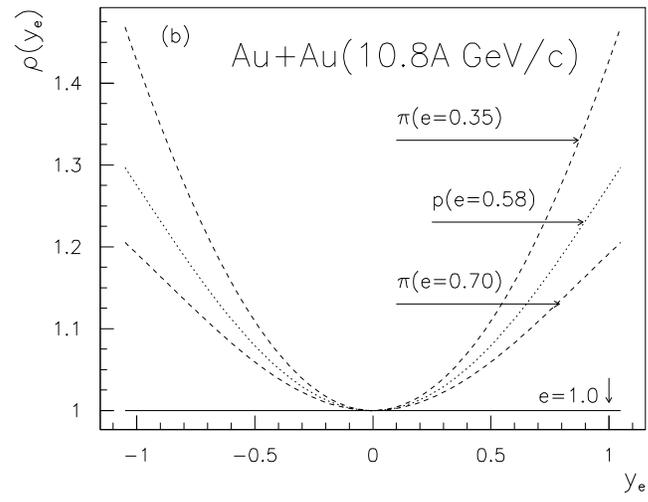
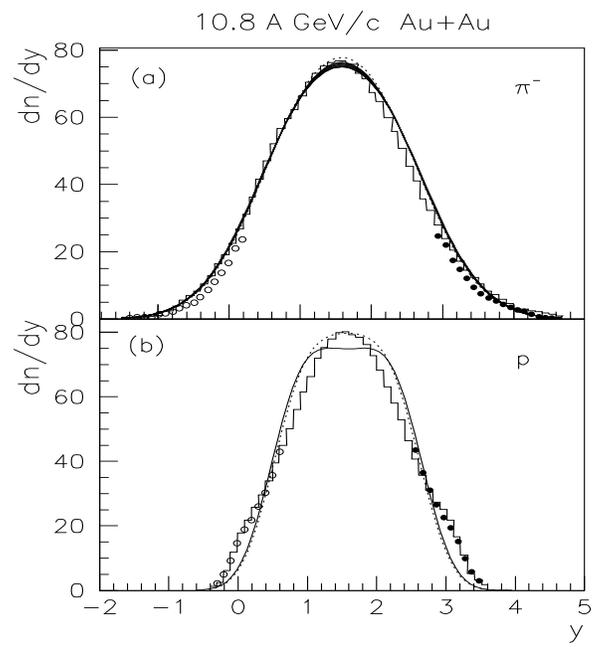
\end{document}